\documentclass[aps,prb,showpacs,twocolumn]{revtex4}

\usepackage{graphicx}
\usepackage{dcolumn}
\usepackage{bm}
\usepackage{times}

\newcommand {\mb}[1]{\mathbf{#1}}

\newcommand {\p}{\phi}
\newcommand {\px}{\phi(\mathbf{x})}
\newcommand {\pk}{\phi(\mathbf{k})}

\newcommand {\sk}{\sum_\mathbf{k}}

\newcommand {\vk}{\mathbf{k}}
\newcommand {\vx}{\mathbf{x}}

\newcommand {\vvr}{\mathbf{r}}

\newcommand {\hs}[1]{\hspace{#1cm}}
\newcommand {\vs}[1]{\vspace{#1cm}}

\newcommand{\be}{\begin{equation}}
\newcommand{\ee}{\end{equation}}
\newcommand{\bea}{\begin{eqnarray}}
\newcommand{\eea}{\end{eqnarray}}
\newcommand{\bc}{\begin{center}}
\newcommand{\ec}{\end{center}}
\newcommand{\bfig}{\begin{figure}}
\newcommand{\efig}{\end{figure}}

\begin{document}



\title{Langevin simulations of a model for ultrathin magnetic films}

\author{Lucas Nicolao}
 \email{nicolao@if.ufrgs.br}
\author{Daniel A. Stariolo}%
 \email{stariolo@if.ufrgs.br}
\affiliation{Departamento de F\'isica,\\ Universidade Federal do Rio Grande do Sul\\ 
CP 15051, 91501-979, Porto Alegre, Brazil}
\altaffiliation{Research Associate of the Abdus Salam International Centre for Theoretical Physics,
Trieste, Italy}

\date{\today}

\begin{abstract}
We show results from simulations of the Langevin dynamics of 
a two-dimensional scalar model with competing interactions for ultrathin magnetic films.
We find a phase transition from a high temperature disordered phase to a low
temperature phase with both translational and orientational orders. Both kinds of order
emerge at the same temperature, probably due to the isotropy of the model Hamiltonian. 
In the low temperature phase orientational correlations show long range order while
translational ones show only quasi-long-range order in a wide temperature range.
The orientational correlation length and the associated susceptibility seem to diverge
with power laws at the transition. While at zero temperature the system exhibits
stripe long range order, as temperature grows we observe the proliferation of different
kinds of topological defects that ultimately drive the system to the disordered phase.
The magnetic structures observed are similar to experimental results on ultrathin
ferromagnetic films.
\end{abstract}

%
%
%

\maketitle

\section{\label{sec:intro} Introduction}

In the last years the interest in understanding the 
thermodynamic and mechanical properties of magnetic ultrathin films has grown considerably.
Part of this interest is obviously motivated by the great amount of
technological applications related to their magnetic behavior, such as data 
storage and electronics~\cite{HuSc1998}.
In recent years the advance in experimental techniques has made possible to study
in great detail the complexity of magnetic order in thin films, where an extremely
 rich phenomenology is present~\cite{PoVaPe2003,WuWoSc2004,Ja2004}.
Part of this phenomenology - as in many other physical and chemical
systems~\cite{SeAn1995} - is due to the presence of competing 
interactions 
acting on different length scales that frustrate the system and leads to 
mesoscopic pattern formation. In the case of an ultrathin magnetic film in
the absence of an external magnetic field, stripe patterns of opposed 
magnetization are formed due to the competition between the ferromagnetic 
exchange interaction and the long-ranged dipolar interaction~\cite{GaDo1982}.

Of particular interest for applications in data storage are films with strong perpendicular 
anisotropy, where spins point preferentially out of the plane. This happens when the 
(perpendicular) anisotropy energy overcomes the effect of dipolar interactions which induce 
an in-plane anisotropy. When this happens the system goes through a ``spin reorientation 
transition'' (SRT)~\cite{AlStBi1990,Po1998,MaDeWh1998}. In this phase, when magnetization points 
preferentially out of plane, complex magnetic structures arise, showing the formation of 
patterns with stripe order \cite{VaStMaPiPoPe2000,PoVaPe2003,WuWoSc2004}. Antiferromagnetic
stripe order dominates the low temperature/low energy behavior. As temperature increases
towards the SRT, perfect stripe order is disrupted by proliferation of topological defects,
like dislocations and disclinations in the magnetic patterns, which eventually drive the
system to the paramagnetic high temperature phase. Understanding how magnetic stripe order
emerges from microscopic interactions and the characterization of the relevant 
thermodynamic behavior in the perpendicular phase is of relevance both from a fundamental 
point of view and for future applications in magnetic memory devices at the nanoscopic level.

Theoretical analysis of thermodynamic behavior rely on elastic Hamiltonian approximations
and show a variety of possible scenarios, strongly
dependent on the behavior of different kind of anisotropies present in the elastic energy
\cite{GaDo1982,AbKaPoSa1995,Po1998}. In \cite{AbKaPoSa1995} two possible scenarios were
anticipated, in one there is a single phase transition from a high temperature paramagnetic
phase to a smectic (stripe) phase at low temperatures. Other possible scenario shows the
presence of an intermediate nematic phase, where translational order is lost but orientational
one persists. Searching for evidence of these scenarios is one of our motivations for the
present work.
There is also a rather large number of numerical simulations and analysis of the different
patterns observed in ultrathin film models. Relevant to the present work are a series of simulations
by MacIsaac and coworkers~\cite{DeMaWh2000}. They made detailed Monte Carlo simulations of 
Heisenberg and Ising  systems with competing exchange and dipolar interactions. The stripe
nature of the ground state in a two dimensional system of Ising spins was determined
analytically in \cite{MaWhRoDe1995}. Phase diagrams showing the presence of the spin
reorientation transition in Heisenberg models with perpendicular anisotropy, dipolar interactions
and with or without exchange interactions were also studied~\cite{MaWhDePo1996,MaDeWh1998}. 
Cannas and coworkers have made a series of detailed predictions on phase diagrams and
dynamic properties of a two dimensional Ising model with competing exchange and dipolar 
interactions~\cite{GlTaCa2002,CaStTa2004,CaMiStTa2006} by Monte Carlo simulations. In
particular, in \cite{CaMiStTa2006} evidence was presented of a new, intermediate, nematic
phase between the stripe and disordered phases. There is also a rather large amount of
numerical work focused on qualitative descriptions of patterns, but few quantitative
approaches.

Simulations of Heisenberg models have been largely  restricted to determinations of phase 
diagrams mainly because analyzing quantitatively the structure of phases and patterns
that emerge is a computationally very hard task in these models. On the other hand, studies
of Ising models have also concentrated on quantitatively determining phase diagrams and
also the presence and evolution of magnetic patterns. Nevertheless in the Ising case the
sharp nature of domain walls induces a rather artificial structure of patterns, where e.g.
tetragonal symmetry dominates completely the scene in square lattices. In this work we
introduce a model intended to  make a bridge between the more realistic but cumbersome
Heisenberg models and the much simpler Ising models which nevertheless are not suitable for 
understanding
the diversity of magnetic structures present in a real ultrathin film. Our results are
qualitative in the sense that we do not intend to reproduce experimental parameter values of
a particular system, but nevertheless
we present a quantitative picture of phase transitions and the behavior of relevant
thermodynamic variables in the regime of perpendicular magnetization, which can be easily
mapped to real situations. Furthermore, the model introduced below admits numerical as
well as analytical
treatment of both equilibrium and dynamical behaviors and a comprehensive picture is
beginning to emerge~\cite{BaSt2007,MuSt2007}.

With this aim, in the present work we introduce and analyze, by means of Langevin simulations, 
a coarse 
grained model of an ultrathin film in two dimensions. By making reasonable approximations 
to the full micromagnetic dynamics, which should be satisfactory in the perpendicular region, 
we address the existence of antiferromagnetic stripe order,  how thermal fluctuations affect 
this order, the possible appearance of a phase transition at finite temperature, and the 
characterization of the magnetic structures relevant in each temperature region. 
Coarse-grained models similar to the one introduced in the next section have been studied in 
the past.  In an early work, Roland and Desai~\cite{RoDe1990} studied the
dynamical process of phase separation and emergence of stripe order in Langevin simulations
of a model for ultrathin films. More recently, Jagla~\cite{Ja2004} explored the different
morphologies and patterns that can appear in such systems, showing a variety of very
interesting phenomena. Furthermore, he showed that the same model is capable of 
reproducing the detailed phenomenology of histeretic behavior known in ultrathin 
films~\cite{Ja2005}.
 Nevertheless, up to our knowledge, more quantitative studies of the 
phase transitions and the different kinds of magnetic order present in these type of models 
have not been addressed up to now.

\section{\label{sec:model} Model and Numerical implementation}

A widely acceptable microscopic description of micromagnetic dynamics is given by the
Landau-Lifshitz-Gilbert equation \cite{HuSc1998}:
\be
\frac{\partial \vec{\phi}}{\partial t} = -\alpha\ \vec{\phi} \times \vec{B} - \gamma\ \vec{\phi}
\times (\vec{\phi} \times \vec{B})
\label{LLG}
\ee
where $\vec{\phi}$ is the three dimensional magnetic moment vector, $\vec B$ is the
effective field acting on it and $\alpha$ and $\gamma$ are microscopic phenomenological
constants. The first term induces a precessional movement of the magnetic
moment around the field, while the second term is a phenomenological one representing
a damping effect which induces the magnetization to align with the field. This
representation of magnetic dynamics is satisfactory in a wide variety of situations but 
it is very difficult to analyze analytically and is also very demanding computationally.
Nevertheless, below the SRT where perpendicular anisotropy dominates, one can obtain a much 
simpler description of magnetization dynamics suitable to our purposes. Considering a single 
moment $\vec{\phi}$ with the effective field $\vec B$ pointing in the  z direction and expanding 
the three components of $\vec{\phi}$ in eq.(\ref{LLG}), one easily finds that the evolution of
the perpendicular component is given by~\cite{Ja2005}: 
\be
\frac{\partial \phi_z}{\partial t} = \gamma \,B\ (1-\phi_z^2)
\ee
where the constraint $\phi_x^2+\phi_y^2+\phi_z^2=1$ was used (this is automatically satisfied by
the LLG dynamics). Then, in the limit of strong perpendicular anisotropy, when the effective
field can be approximated to point preferentially along the z direction, the evolution of the 
perpendicular component is approximately autonomous, although the other two components are not. 
Consequently from now on we will restrict the analysis to the z component and drop out the
corresponding subindex.

In the system of interest, the effective field will be composed typically of three contributions
of the form:
\be
B = h + a\ \phi - b \int d^{2}\mb{x'}\ J(|\mb{x}-\mb{x'}|)\ \p(\mb{x'})
\label{eff.field}
\ee
where $h$ is an external field, the second term is a perpendicular anisotropy field and the
last term will be the dipolar field in our case. The constants can be easily related with
experimental ones, but this is not necessary for the purposes of this work, which is to study
some universal properties of the phases and patterns emerging in such a system.



The complete energy function of the model consists of both a 
local and a nonlocal terms:
\begin{equation}
H[\p]=H_L[\p]+H_{NL}[\p]
\label{H1}
\end{equation}

According to (\ref{eff.field}) for the case of zero external field,  the local potential can 
be written in full generality:
\be
H_L[\p]=\frac{1}{2}\int d^{2}\mb{x}
\big[\kappa(\boldsymbol{\nabla}\p(\mb{x}))^{2}
-r\p^2(\mb{x})+\frac{u}{2}\p^4(\mb{x}) \big]
\label{H_L}
\ee
In order that the local potential assume the desired 
double-well structure, $u$ and $r$ are taken to be positive phenomenological 
constants. We have also added a continuous approximation for the exchange interaction in the
form of an attractive square gradient term, which favors spatial homogeneity of 
the order parameter. 

Disregarding higher order interactions in (\ref{eff.field}), the nonlocal term modeling a 
repulsive dipolar interaction has the form:
\begin{equation}
H_{NL}[\p] = \frac{1}{2\delta}
\int d^{2}\mb{x}\int d^{2}\mb{x'} \p(\mb{x})J(|\mb{x}-\mb{x'}|)\p(\mb{x'})
\label{H_NL}
\end{equation}

where $J(|\mb{x}-\mb{x'}|)=1/|\mb{x}-\mb{x'}|^3$ in two dimensions.
In equations (\ref{H_L}) and (\ref{H_NL}), $\kappa$ and $\delta$
are positive phenomenological constants that describe, respectively, the 
range of the short-range exchange interaction and the strength of the long-range dipolar one.
In all space integrations it is assumed a short distance cutoff $2\pi /\Lambda$,
which in the numerical implementation appears naturally as the lattice constant $a$.

Hence, in the limit of strong perpendicular anisotropy and strong damping, the relaxational 
dynamics of the system 
can be modeled by the following Langevin equations:
\be
\label{langevin1}
\frac{{\partial\p(\mb{x},t)}}{\partial t}=-\Gamma\frac{{\delta H[\phi]}}{\delta\p(\mb{x},t)} + \eta(\mb{x},t)
\ee
where $\Gamma$ is a constant (the mobility) which sets the time scale,  and $\eta(\mb{x},t)$ is a 
Gaussian thermal noise with $\left\langle \eta(\mathbf{x},t)\right\rangle=0$ and 
$\left\langle \eta(\mathbf{x},t)\eta(\mathbf{x}',t')\right\rangle=2\,\Gamma\,T\,\delta(t-t')\,\delta^2(\mathbf{x}-
\mathbf{x'})$, as usual. Equation (\ref{langevin1}) can be 
explicitly written as:
\bea
\frac{1}{\Gamma}\frac{{\partial\phi(\mathbf{x},t)}}{\partial t}&=&
\kappa\,\nabla^{2}\phi(\mathbf{x},t)+r\phi(\mathbf{x},t)-u\phi^3(\mathbf{x},t)\label{L1}\\
&-&\frac{{1}}{\delta}\int d^2\mathbf{x'}\phi(\mathbf{x'},t)J(|\mathbf{x}-\mathbf{x'}|)+\eta(\mathbf{x},t)/\Gamma \nonumber
\eea

It is convenient to express the above equations in 
dimensionless form by means of a transformation of variables. 
Following Roland and Desai \cite{RoDe1990}, 
this transformation leads to the dimensionless form of equation (\ref{L1}):
\bea
\frac{{\partial\phi(\mathbf{x},t)}}{\partial t}&=&\nabla^{2}\phi(\mathbf{x},t)+\phi(\mathbf{x},t)-\phi^3(\mathbf{x},t) \nonumber\\
&-&\frac{{1}}{\delta}\int d^2\mathbf{x'}\phi(\mathbf{x'},t)J(|\mathbf{x}-\mathbf{x'}|)+\eta(\mathbf{x},t)\label{Ladim}
\eea
with a short distance cutoff $2\pi/\Lambda$ and thermal noise correlation 
$\left\langle \eta(\mathbf{x},t)\eta(\mathbf{x}',t')\right\rangle=2 T\,
\delta(t-t')\,\delta^2(\mathbf{x}-\mathbf{x'})$. 
Now the parameter $\delta$ stands for the relative strength between the two competing interactions.
The last Langevin equation can be written in Fourier space as:
\be
\frac{\partial\p(\vk,t)}{\partial t}=
-A(k)\p(\vk,t) +\left[-\p^3(\vx,t) +\eta(\vx,t)\right]_{\vk}^F \label{Lk}
\ee

To be compatible with the subsequent numerical implementation of this 
equation, the last two terms were not explicitly written in Fourier space. 
Here $\big]_{\vk}^F$ means the ${\vk}$ component of the
corresponding Fourier transform.
The function $A(k)$ corresponds to:
\be
A(k)=k^2-1+J(k)/\delta \label{Ak}
\ee
which encodes all spatial information about the interactions.
If this quantity has a negative minimum at a wave vector $k_m$, 
selected by varying $\delta$, the solution
of equation (\ref{Lk}) is a modulation in a single direction 
with periodicity given by $k_m$. 




We have numerically solved the stochastic differential equation 
(\ref{Lk}) discretizing space in a square lattice with mesh size $a$.
Periodic boundary conditions were implemented using the Ewald summation 
technique \cite{FrSm1996} in the long-range dipolar interaction.
After spatial discretization this interaction is no longer isotropic for
all spatial scales and it becomes gradually anisotropic as the wave vector 
comes close to $\pi$. 
At the relevant spatial scales for our simulations, $J(\vk)$ is 
slightly anisotropic. It is important to note that symmetry properties of
the different magnetic structures appearing at low temperatures are affected
by the square symmetry of the lattice. In a triangular lattice, for example,
the phenomenology may be to some extent different~\cite{Ve1998}.

We found it advantageous to use a spectral method, since in the
Fourier space form of the Langevin equation (\ref{Lk})
both spatial derivatives and the dipolar interactions acquire an 
algebraic form.

The time derivative was approximated using a simple Euler scheme with
a time step $\Delta t$. 
Taking an isotropic form of the discretized laplacian, 
the spatial derivatives were treated using a semi-implicit method, where 
the $k^2$ term in (\ref{Lk}) is evaluated in the new time value. 
This treatment is standard to improve the stability of the algorithm 
\cite{PrTeVeFl1992}.

Therefore, discretizing equation (\ref{Lk}) in such manner, i.e.~through a
first order semi-implicit spectral method, we obtain the following recurrence 
relation:
\bea
\p(\vk,t+\Delta t) &=& \frac{1}{1+\Delta t\, k^2}\hs{0}\{
\left[1+\Delta t-\Delta t\,a^2 J(\vk)/\delta \right]\p(\vk,t) \nonumber \\ 
&&+\left[-\Delta t\,\p^{3}(\vx,t)+
\eta(\vx,t)\right]_{\vk}^{F}\,\}
\label{Ld}
\eea


After discretization the noise term $\eta$, in the way it appears in the last
equation,
is a random Gaussian number with amplitude $\left(2T\Delta t/a^2\right)^{1/2}$.
The dipolar interaction $J(\vk)$ in (\ref{Ld}) is the fast Fourier transform
of the result of the Ewald summation, evaluated 
at the beginning of the simulation.

The computational advantage of updating the system in Fourier 
space is 
accomplished using an 
adaptative FFT algorithm \cite{FrJo2005}, where the main time consuming operations 
are transforming Fourier the field $\p$ and the $(-\p^3 + \eta)$~term and then 
transforming back the new field value.

\section{Results}

We performed simulations of the continuum dipolar model through equation 
(\ref{Ld}).~We analyzed the case 
$\delta=2 \sqrt 5$, 
where the ground state is 
a stripe modulated state with wave vector $k_m=1.01447$, which is close to $\pi/3$.
Simulations were performed for system sizes $L=192$ and $L=384$ for different temperatures.

The existence of a nontrivial solution with wave vector $k_m$
of the Langevin equation (\ref{Lk}) at zero temperature
depends on whether the term in (\ref{Ak}) is negative
at $k=k_m$.
Before the variables transformation, 
this could be done
varying the parameter $r$. After the transformation, the existence
of a solution is achieved only by tuning the mesh size $a$.
We set $a=\sqrt 5$. The time 
integration was stable inside the range of temperatures of
the simulations using a time step $\Delta t=0.5$.

To measure the orientational order of the stripes we first consider the
director field:
\be
\hat{n}(\mathbf{x})=\frac{\boldsymbol {\nabla }\phi (\vx)}{|\boldsymbol {\nabla }\phi (\vx)|}
\ee
which provides the local orientation of the stripe, when localized on a domain 
wall between opposite magnetizations.
An analogy between this 
quantity and the Frank director 
of smectic liquid crystals has already
been made on previous works 
\cite{ToNe1981,ChBr1998,QiMa2003}, and leads to the definition
of a tensor order parameter:
\be
Q_{\alpha \beta}(\vx)= n_{\alpha }(\vx)n_{\beta }(\vx)-\frac{1}{2}\,\delta_{\alpha \beta}
\label{qotensor}
\ee
where $\alpha ,\beta =1,2$ are the cartesian components.
Similar to a nematic order parameter in a liquid crystal, 
an orientational order parameter $Q$ can be defined as the positive eigenvalue
of the spatial average of the above tensor order parameter~\cite{CuFr1990}. 
This value corresponds to the quantity $\cos 2\theta$, where $\hat{n}=(\cos \theta,
\sin \theta)$ and $\theta$ is the angle between the local director field and the
mean orientation of the system.
But, since the director field inside the stripe domains does not necessarily
provide the local orientation of the stripe, to get a more precise value for
the orientational order parameter we average Eq.(\ref{qotensor})
only over domain wall sites, namely:
\be
\overline{Q}_{\alpha \beta} =\frac{1}{L^\prime}\sum_{\vx}\!^\prime\,Q_{\alpha \beta}(\vx)
\ee
where the prime denotes the restricted sum with $L'$ being the total number of 
domain wall sites. Now we can write explicitly the orientational order 
parameter as:
\be
Q=\sqrt{\overline{Q}_{11}^{\>2} + \overline{Q}_{12}^{\>2}}
\label{qo}
\ee

One possible way to characterize the translational order of the 
stripes is through a staggered magnetization, defined as:
\be
m_{\mb{k}}=\left< \frac{1}{L^2} \sum_{\vx}  sgn\left(\phi(\vx)\right)
sgn\left(\cos(\mb{k}\cdot\vx)\right) \right>
\label{mstaggered}
\ee
where $sng$ is the sign function and for $\mb{k}$ we use the ground state 
wave vector.

The results we present here were obtained with the
following procedure: the system is initialized in the ground state
and heated with a heating rate $d_{T}$. At each temperature of interest, the
heating process is halted and the system is left to
run a transient period of $n_t$ time steps before we start recording
system configurations at each $n_m$ time steps, used later to measure
the desired quantities. Typical values of $n_t$ were between $5\times10^4$ to
$4\times10^6$ in order to get as close as possible to equilibrium.
%
To estimate both transient ($n_t$) and decorrelation ($n_m$)
times we analyzed the behavior of the two-times correlation
function for the different system sizes and temperatures.
%
Typical $n_m$ values were in the order of $10^5$ and the results are averages
over 20 to 60 system configurations.

\begin{figure}[ht!]
\includegraphics[width=8.7cm]{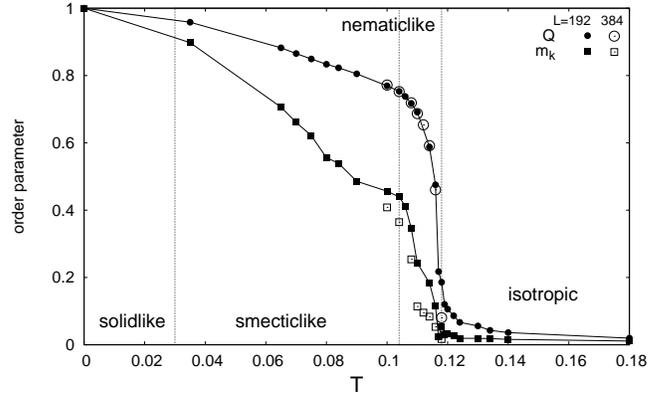}
\caption{Order parameters as a function of temperature for $L=192$.
The open symbols corresponds to the larger system size, $L=384$.
Thin vertical lines separate approximately temperature ranges
of the different regimes.}
\label{pmos}
\end{figure}

\begin{figure}[t]
(a) \hs{3.9} (b)\\
\includegraphics[scale=.64]{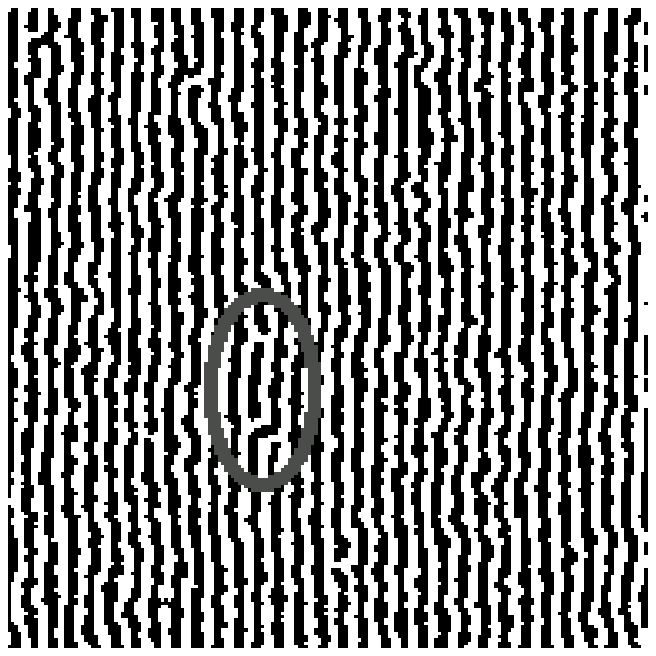}~\includegraphics[scale=.64]{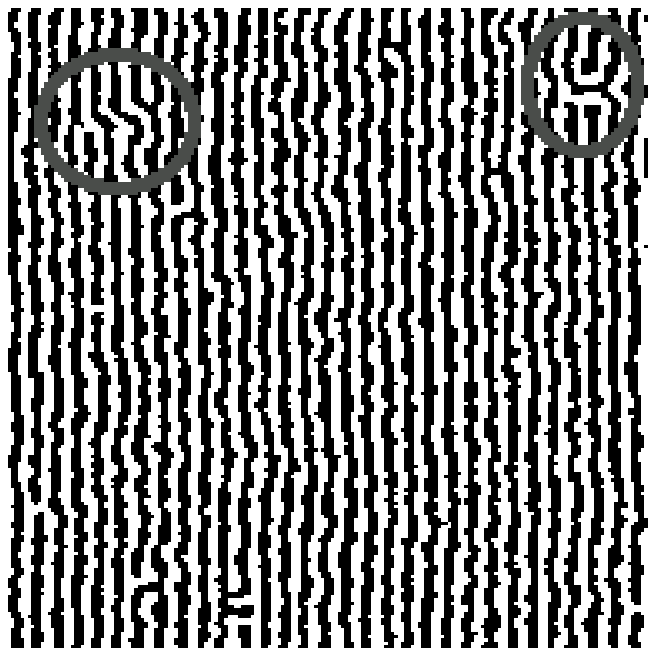}
\caption{Typical configurations in the smectic phase for $L=192$
at (a) $T=0.1$ and (b) $T=0.104$. In (a) the encircled region exhibits two 
pairs of dislocations; above there is a "passage" due to two dislocations 
inserted in the same stripe, and below a pair of dislocations separated by one 
stripe. In (b) a pair separated by four stripes and a pair formed by a larger Burgers vector dislocation and a double dislocation.}
\label{typical0.1}
\end{figure}

In Fig.~\ref{pmos} we show the results for the translational and orientational order parameters.
The data clearly shows a phase transition, where both kinds of order emerge at a finite
temperature.
In the $L=192$ case we found that the two order parameters 
decay simultaneously to zero at $T\approx0.12$, even 
though the staggered magnetization drops significantly at $T=0.11$. In the region in
between the translational order parameter seems to present strong finite size effects.
As the temperature is lowered in the ordered phase, a stripe structure develops
gradually by the annihilation of defects (see Fig.~\ref{typical0.1}).
Although we were unable to study in more detail the finite size scaling near the
transition, it seems improbable that an intermediate, nematic-like phase, with only
orientational order can be present in this model. Although translational order decays 
faster than the orientational one, our results point to the presence of a single phase
transition from a disordered high temperature phase to a low temperature phase with
both orientational and translational orders. This is one of the possible scenarios
that emerge from a theoretical analysis of a similar model by Abanov et al.
~\cite{AbKaPoSa1995}.

\begin{figure}[t]
\includegraphics[width=8.7cm]{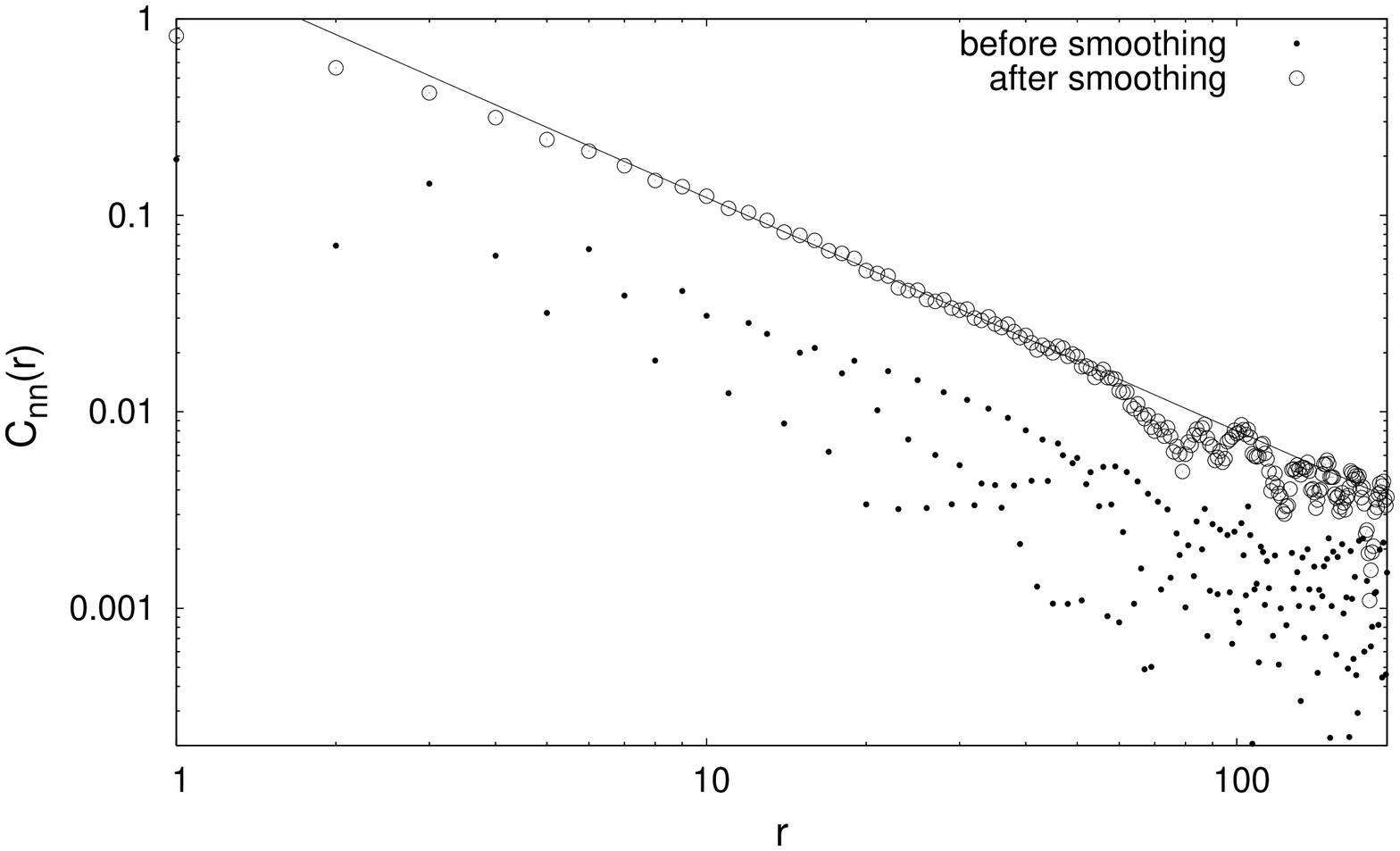}
\includegraphics[width=8.7cm]{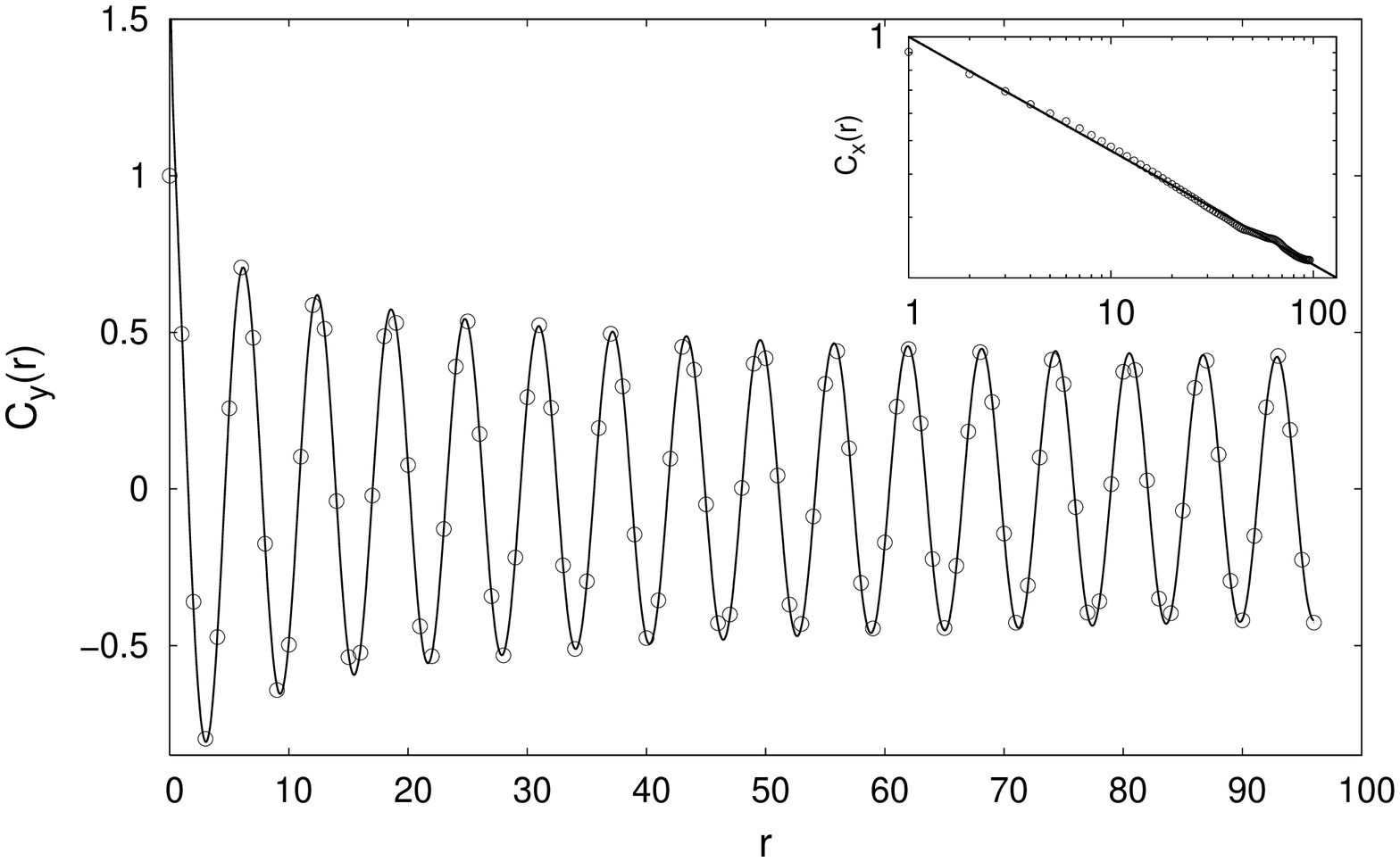}
\caption{Correlation functions in the smectic-like region at $T=0.1$.
The upper figure shows the connected orientational correlation function 
in the perpendicular direction for $L=384$, plotted in log-log scale, 
before and after the smoothing procedure. 
The continuous line corresponds to a fitting by the
function $Ar^{-\alpha_{y}}$, with $\alpha_{y}=1.18$ and $A$ a fitting parameter. 
In the parallel direction
this function decays immediately to zero. The lower figure shows the 
parallel (inset) and perpendicular spatial correlations functions, 
both for $L=192$. The parallel $C_x$ function is plotted in a log-log scale.
The continuous lines corresponds to fittings by the
functions $r^{-\omega_{x}}$ and $\cos(k_0 r)r^{-\omega_{y}}$
with $\omega_{x}=0.33$ and $\omega_{y}=0.19$ and $k_0=1.014$. 
Similar values where found in the $L=384$ case.}
\label{corr0.1}
\end{figure}

\begin{figure}[ht]
\includegraphics[width=8.7cm]{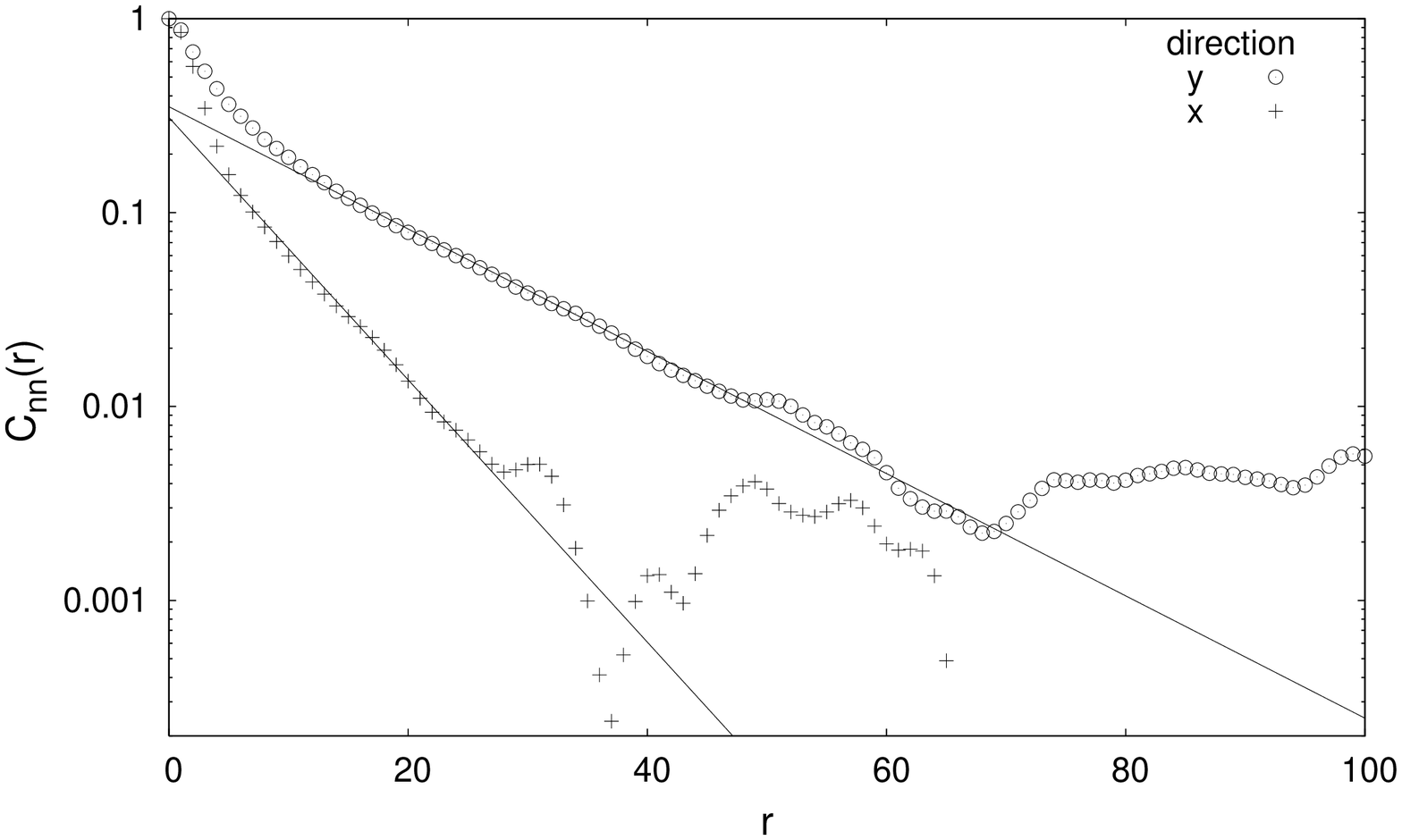}
\includegraphics[width=8.7cm]{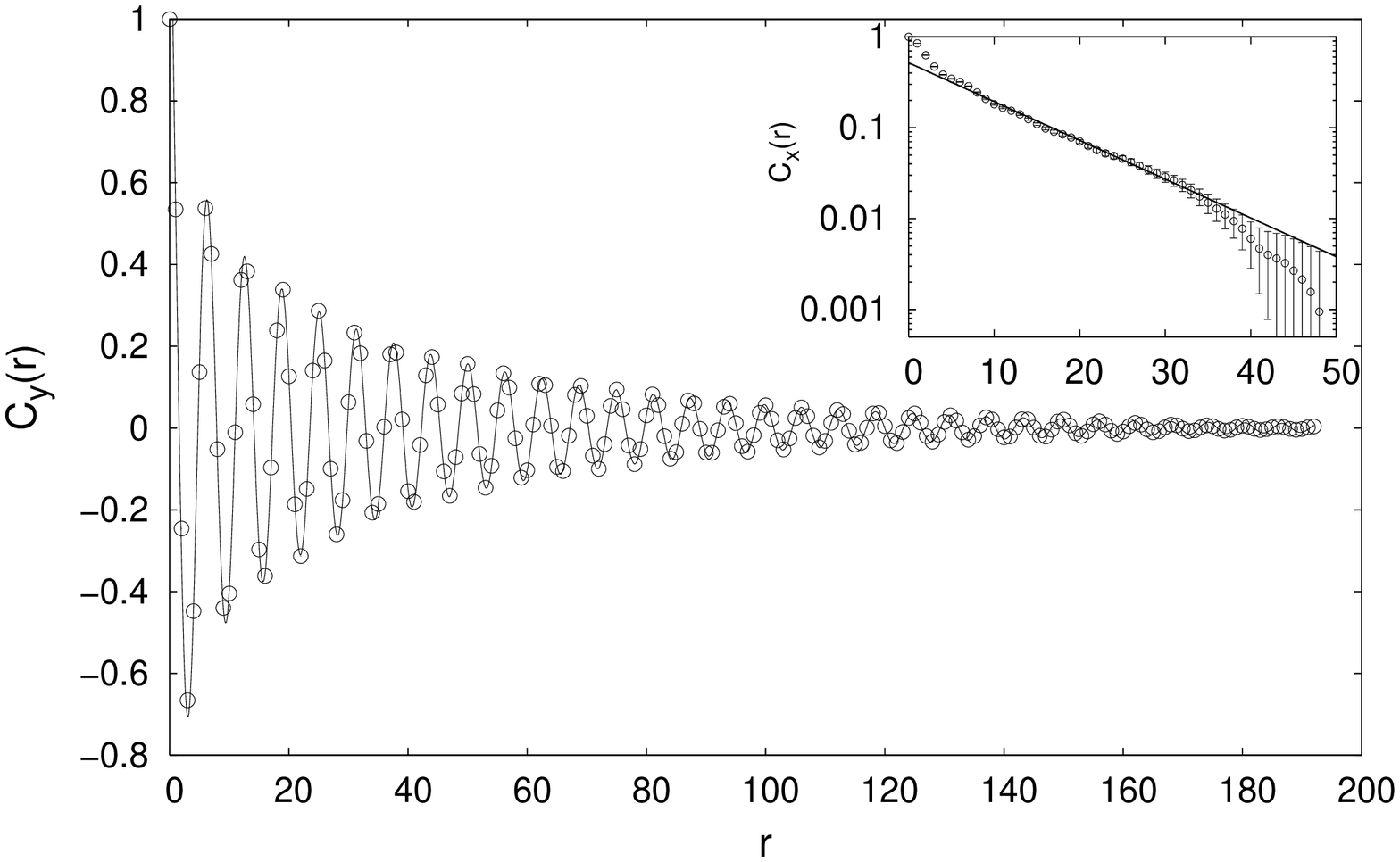}
\caption{Correlation functions in the nematic-like region at $T=0.114$ and $L=384$.
The upper figure shows the connected orientational correlation function 
in the perpendicular and parallel directions in a log-linear scale. 
The continuous lines corresponds to fittings by the
function $A\exp(-r/\lambda)$, with $\lambda_{y}=13.77$ and 
$\lambda_{x}=6.42$.
The lower figure shows the 
parallel (inset) and perpendicular spatial correlations functions.
The parallel $C_x$ function is plotted in a log-linear scale.
The continuous lines 
correspond to fittings by the functions $A\exp(-r/\xi_x)$ and
$\cos(k_0 r)\exp(-r/\xi_y)r^{-\omega_{y}}$ with $\xi_x=10.17$ and 
$\xi_y=60.0$ and $\omega_{y}=0.26$ and $k_0=0.98$.}
\label{corr0.114}
\end{figure}

To take a closer look at the structural properties of the different
magnetic configurations, we calculated the static structure factor:
\be
S(\vk)\equiv\left< \left| \pk \right|^2 \right>
\label{defSk}
\ee
and the associated spatial correlation function, given by:
\be
C(\vvr)=\frac{1}{L^{2}}\sk e^{i\vvr \cdot \vk}S(\vk),
\label{defCr}
\ee
that can be quickly computed using a fast Fourier transform \cite{FrJo2005}. 
The relevant directions for the spatial correlation function are the 
directions parallel and perpendicular to the stripes, respectively denoted 
by $C_x$ and $C_y$. This quantities describe the translational order along the
two relevant directions. We have also computed
nematic (or orientational) correlation functions \cite{ChBr1998,QiMa2003}, 
since they encode information on the 
spatial decay of orientational order of the stripes. Nematic correlations are defined as:
\bea
C_{nn}(\vvr)&=&\frac{2}{L^2}\sum_{\vx}\left<\mathrm{Tr}\,\,Q(\vx+\vvr)Q(\vx)\right> \label{cnntr}\\
&=&\frac{1}{L^2}\sum_{\vx} \left< Q_{11}(\vx+\vvr)Q_{11}(\vx)\right.\nonumber\\
& &+ \left. Q_{12}(\vx+\vvr)Q_{12}(\vx)\right> \label{cnn}
\eea
where the function summed up in Eq.~(\ref{cnntr}) is analogous to 
$\left< \cos\left[ 2\theta(\vx+\vvr)-2\theta(\vx) \right]\right>$.
Examples of translational and orientational correlation functions can be seen
in Figs.~\ref{corr0.1} and \ref{corr0.114} for two different temperatures.

For low enough temperatures ($T<0.11$), the nematic correlations are strongly
affected by residual oscillations of the field $\px$. In order to obtain 
accurate values for the orientational correlation function 
we found convenient to smooth the tensor order parameter 
(\ref{qotensor}) following a smoothing procedure introduced in Ref.\cite{QiMa2003}.
We smooth the fields $Q_{11}(\vx)$ and $Q_{12}(\vx)$ using for each the iterative process:
\be
f_{(n+1)}(\vx)=\frac{1}{2}f_{(n)}(\vx)+\frac{1}{8}\sum_{\vx' \in NN}f_{(n)}(\vx')
\label{smooth}
\ee
where $f_{(n)}$ is one of the fields after $n$ iterations, and $NN$
means the four nearest neighbors of $\vx$ on the square lattice. We found that 
three iterations are enough to get sensible results (see Fig.\ref{corr0.1}). 
Similar smoothing procedures for orientational correlation functions were used
in simulations of melting of two-dimensional solid systems \cite{LiZhTr2006,Ja1999}.
For clarity, in this work we always show the connected orientational correlation function,
where the mean square orientational order parameter is substracted from Eq.(\ref{cnn}).
In this way, $C_{nn}$ accounts for direction fluctuations only.

At sufficiently low temperatures ($T<0.03$) spatial correlations decay
immediately to a constant and there is translational and orientational long-range
order (see Fig.~\ref{pmos} for the values of the order parameters in the 
corresponding temperature regimes). 

At slightly higher temperature and in a wide range ($0.03<T<0.11$) we found 
evidence of a low temperature smectic-like regime, similar to 
the smectic crystal phase predicted by Abanov et al.~\cite{AbKaPoSa1995} 
 and observed experimentally by Portmann et al.~\cite{PoVaPe2003}.
In this phase, undulation fluctuations (meandering excitations)
alone are sufficient 
to cause algebraic decay of translational correlations at low temperatures for $T>0.03$. But at $T=0.09$
bound pairs of dislocations appear and become more common as temperature 
increases. Long-range {\em orientational} order persists over a wider range of temperatures, but
above $T=0.065$ orientational order also starts to decay algebraically.

An example of the algebraic behavior of the correlation functions in 
this region is shown in Fig.~\ref{corr0.1} for $T=0.1$. It is important
to note that the exponent of the translational algebraic decay in this regime 
increases with temperature and is different in the perpendicular and parallel 
directions - exponents values range from $\omega_{x}=0.034$ and 
$\omega_{y}=0.036$ for $T=0.035$ to $\omega_{x}=0.365$ and 
$\omega_{y}=0.212$ for $T=0.104$.
As for the orientational order we observe a small temperature dependence of the 
algebraic decay exponent in the perpendicular direction, with exponents lying
in the range $1.18 < \alpha < 1.43$. In the parallel direction, the
orientational correlation function is a constant, indicating long-range
order in this direction. 

In Fig.~\ref{typical0.1} we illustrate some typical configurations of this phase. 
We see that it is characterized by undulation excitations and a finite density of dislocation 
pairs - some different types are shown encircled.

In a narrow range of temperatures, $0.11 < T < 0.118$, were the orientational order parameter 
is still high but the staggered magnetization drops
considerably fast, we found a change  of behavior in the
correlation functions.
In the parallel direction, translational order is decorrelated exponentially rapidly to zero, 
indicating absence of translational order of the stripes.
In the perpendicular direction, where order is more robust, 
there is an intermediate kind of behavior where the translational correlation 
function is better fitted by a product of a power law and an exponential. 
An example of this behavior is shown in Fig.~\ref{corr0.114}, where one can also 
see that the connected orientational correlation functions now decay exponentially
in both directions but with different correlation lengths. Orientational 
domains in this regime are larger in the perpendicular direction.

\begin{figure}[ht]
(a)\\
\includegraphics[scale=.64]{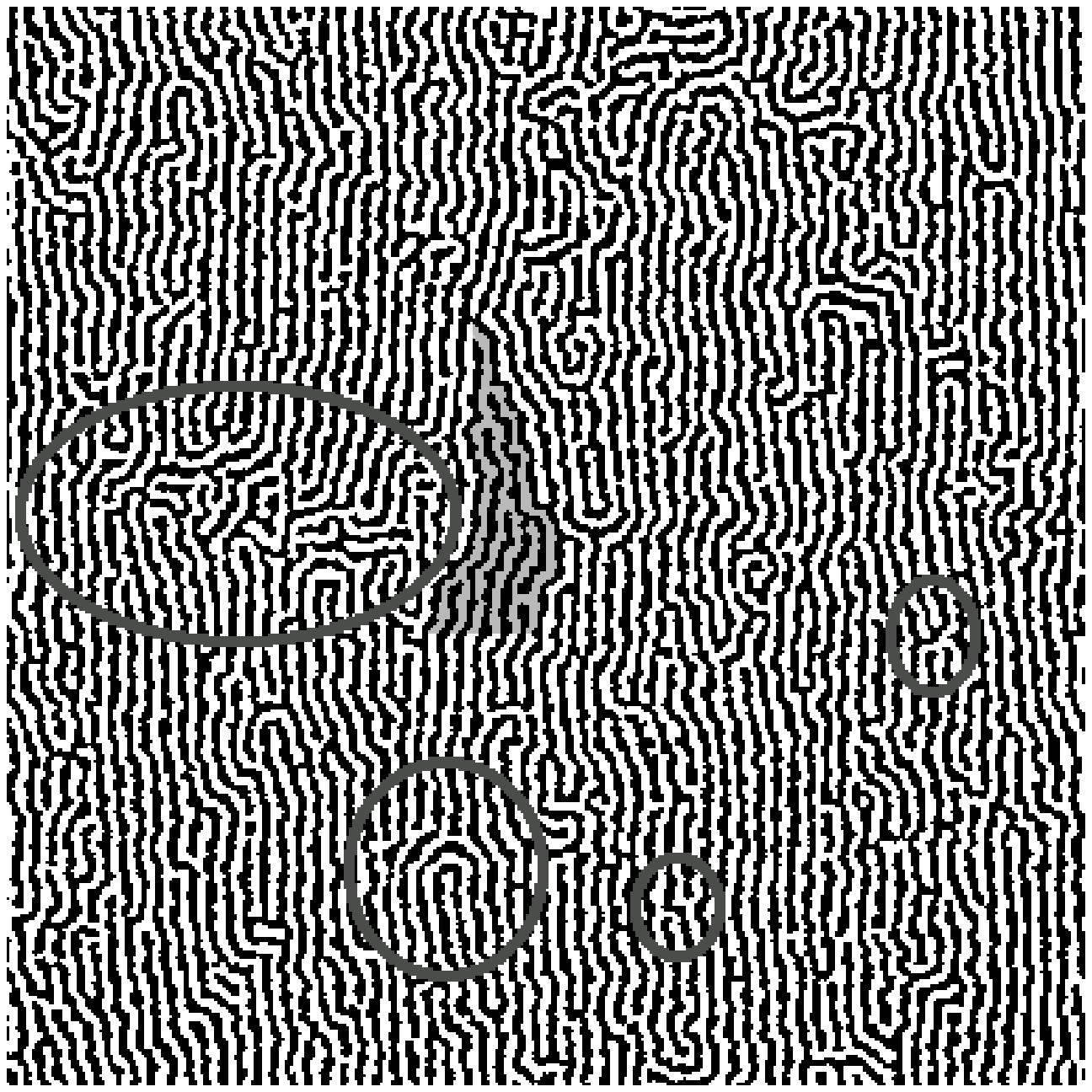}\\
(b)\\
\includegraphics[scale=.75]{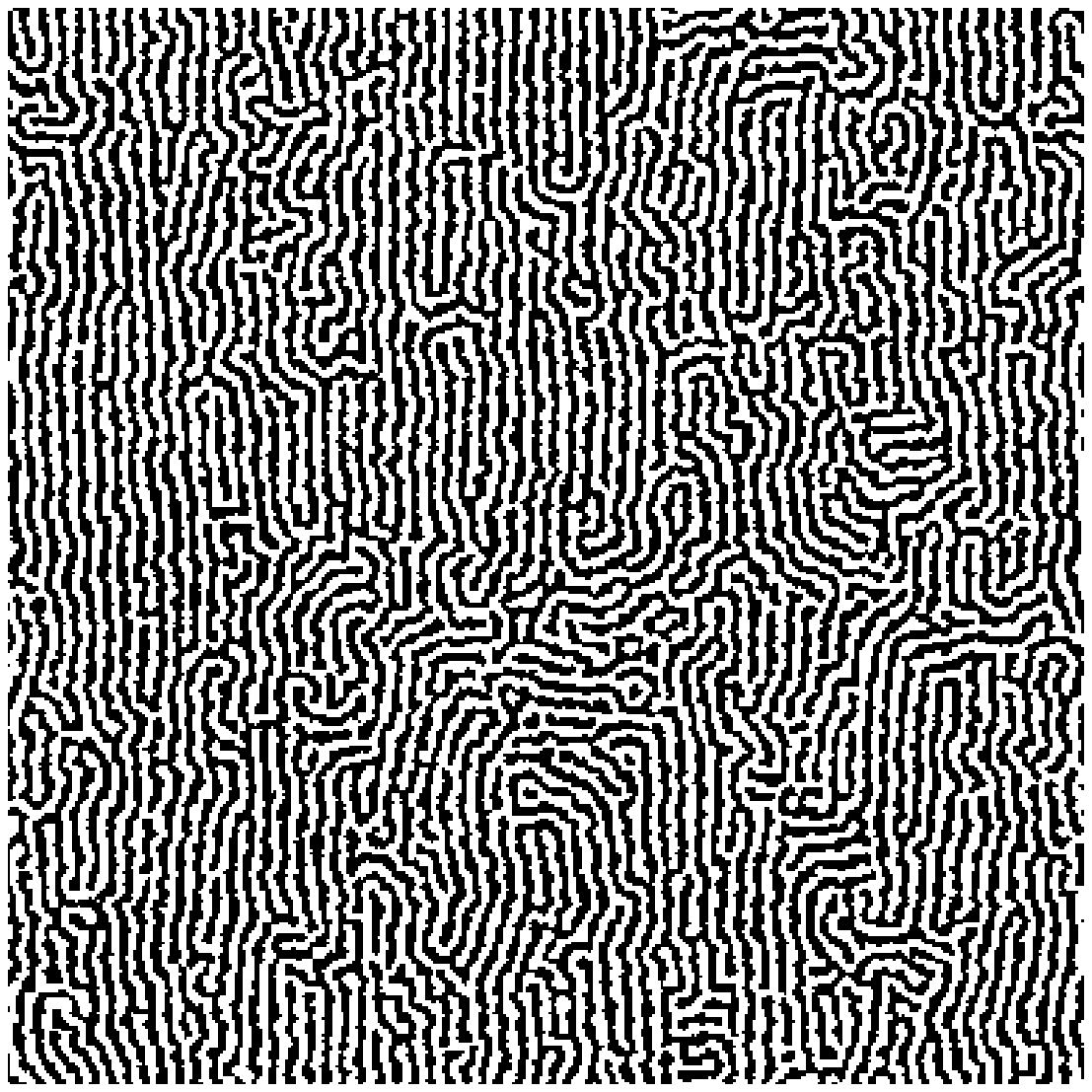}
\caption{Typical configuration in the nematic-like region at (a) $T=0.114$ and (b)
  $T=0.116$ for $L=384$. In (a) some defects are put in evidence. In small circles,
  two pairs of dislocations inside smectic-like domains; in the larger circle a
  disclination dipole; shown colored a series of dislocations inside a single 
  stripe and in an ellipse a domain of perpendicular orientation.}
\label{typical0.114}
\end{figure}

Analyzing visually the configurations of this region 
we observe many excitations disrupting orientational order. A typical 
configuration is shown in Fig.~\ref{typical0.114}. We see that this regime presents large 
Burgers vector dislocations that can be
regarded as a tightly-bound pair of oppositely charged disclinations
\cite{ToNe1981}. There are also what may be called a dislocation cascade, a
series of bifurcations within a single stripe (colored in Fig.~\ref{typical0.114}a). Small
domains of perpendicular orientation (encircled in Fig.(\ref{typical0.114}a)) are present as
well, and become more common and larger as temperature increases (see Fig.~\ref{typical0.114}b).
Surrounded by topological defects there are domains of locally
smectic-like arranged stripes, where order is 
decorrelated mainly by meandering excitations 
and all kinds of pairs of dislocations (some are encircled in Fig.~\ref{typical0.114}a).

\begin{figure}[t]
\includegraphics[width=8.7cm]{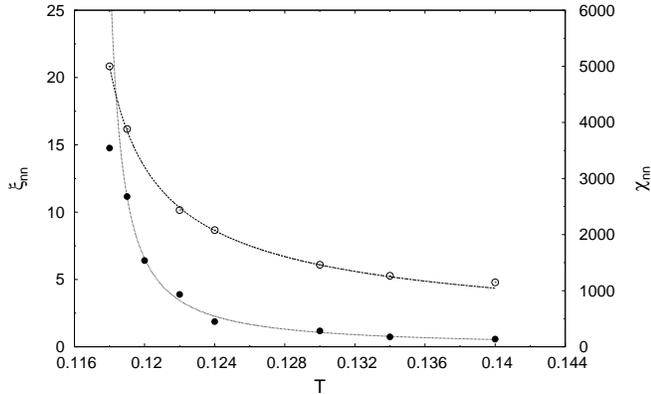}
\caption{Orientational correlation length (full symbols) and susceptibility 
(open symbols) as a function of temperature in the isotropic phase. The
continuous curves show power-law fittings, namely: 
$\xi_{nn}(T)\sim(T-0.1172)^{-1.19}$ and $\chi_{nn}(T)\sim(T-0.1161)^{-0.61}$.}
\label{xichi}
\end{figure}

In order to estimate the transition temperature $T_c$ between the orientationally ordered and
 isotropic phases, 
we have measured the orientational
correlation length and susceptibility of $Q$ in the isotropic phase. 
The results  for $L=192$ are shown in Fig.~\ref{xichi} , together with power-law
fits.
We found that $T_c$ lies close to $T_c \approx 0.117$ and that divergences
are well fitted by power-law forms, at variance with the
exponential divergence that one would expect in a defect-mediated transition
according to the KTHNY theory \cite{ToNe1981,LiZhTr2006}. The power law behavior seems
to be in agreement with a recent result in which the nematic transition is predicted to
be second order~\cite{BaSt2007}.

\begin{figure}[ht!]
\includegraphics[width=8cm]{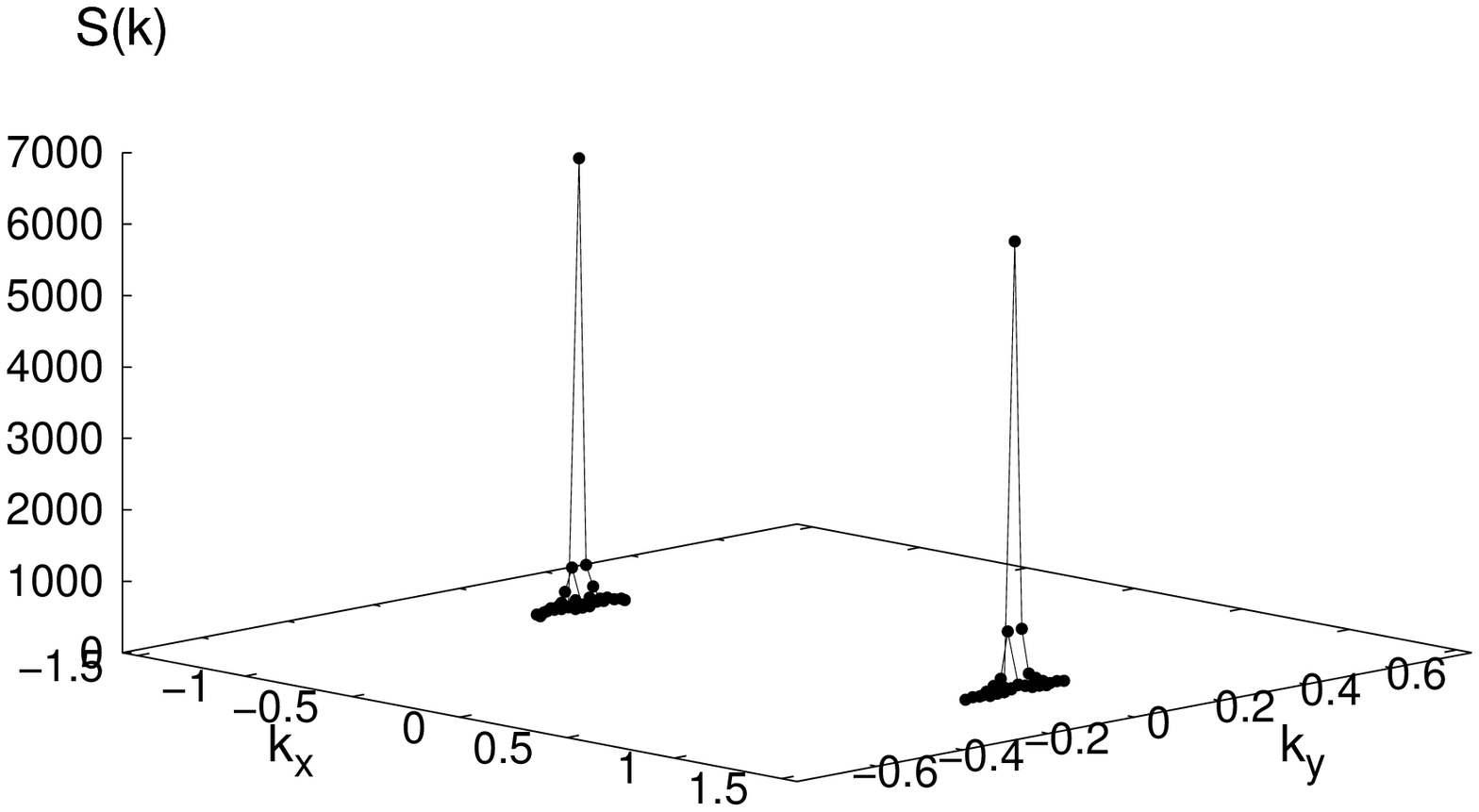}\\
\vs{-1.3}
\includegraphics[width=8cm]{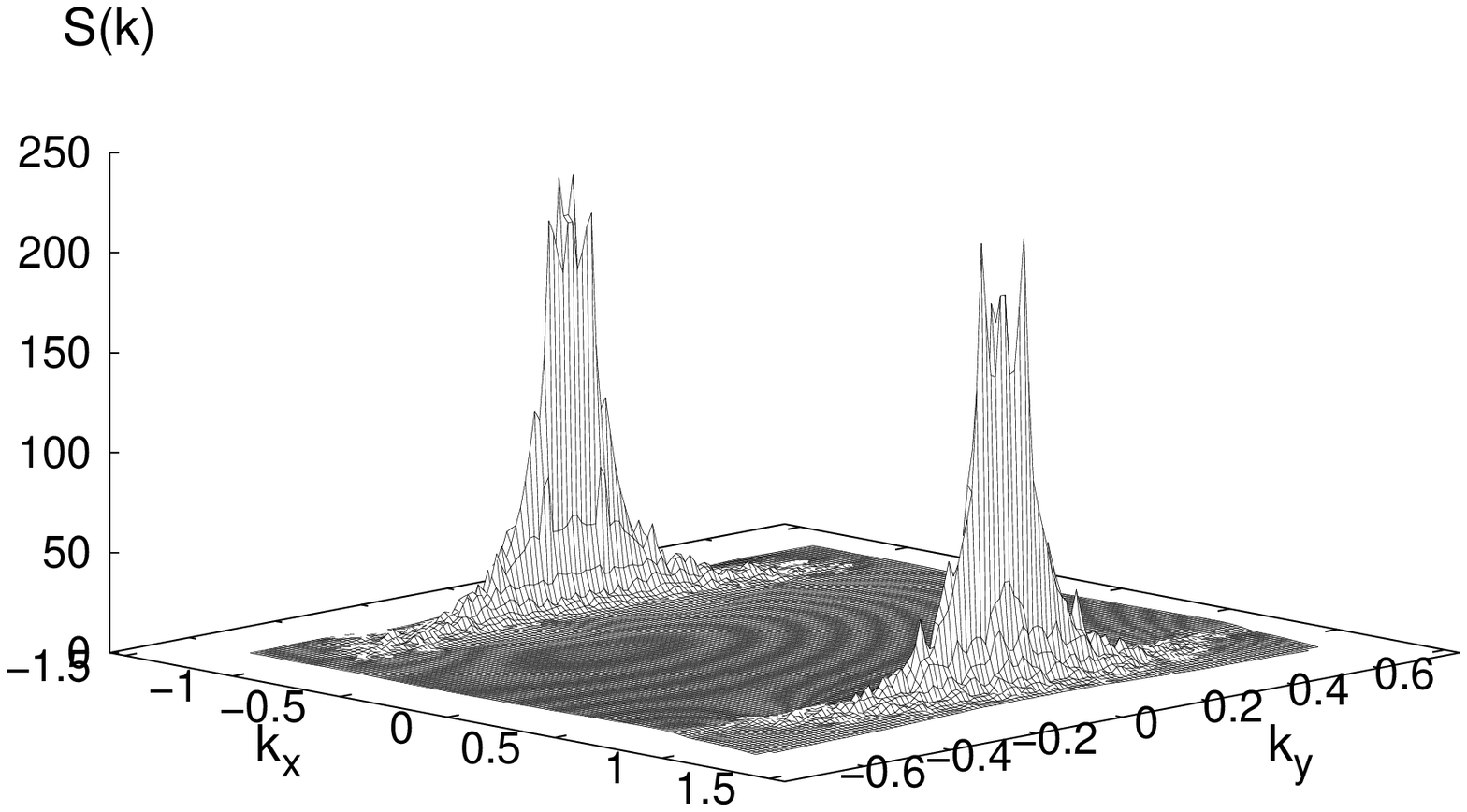}\\
\vs{-1.3}
\includegraphics[width=8cm]{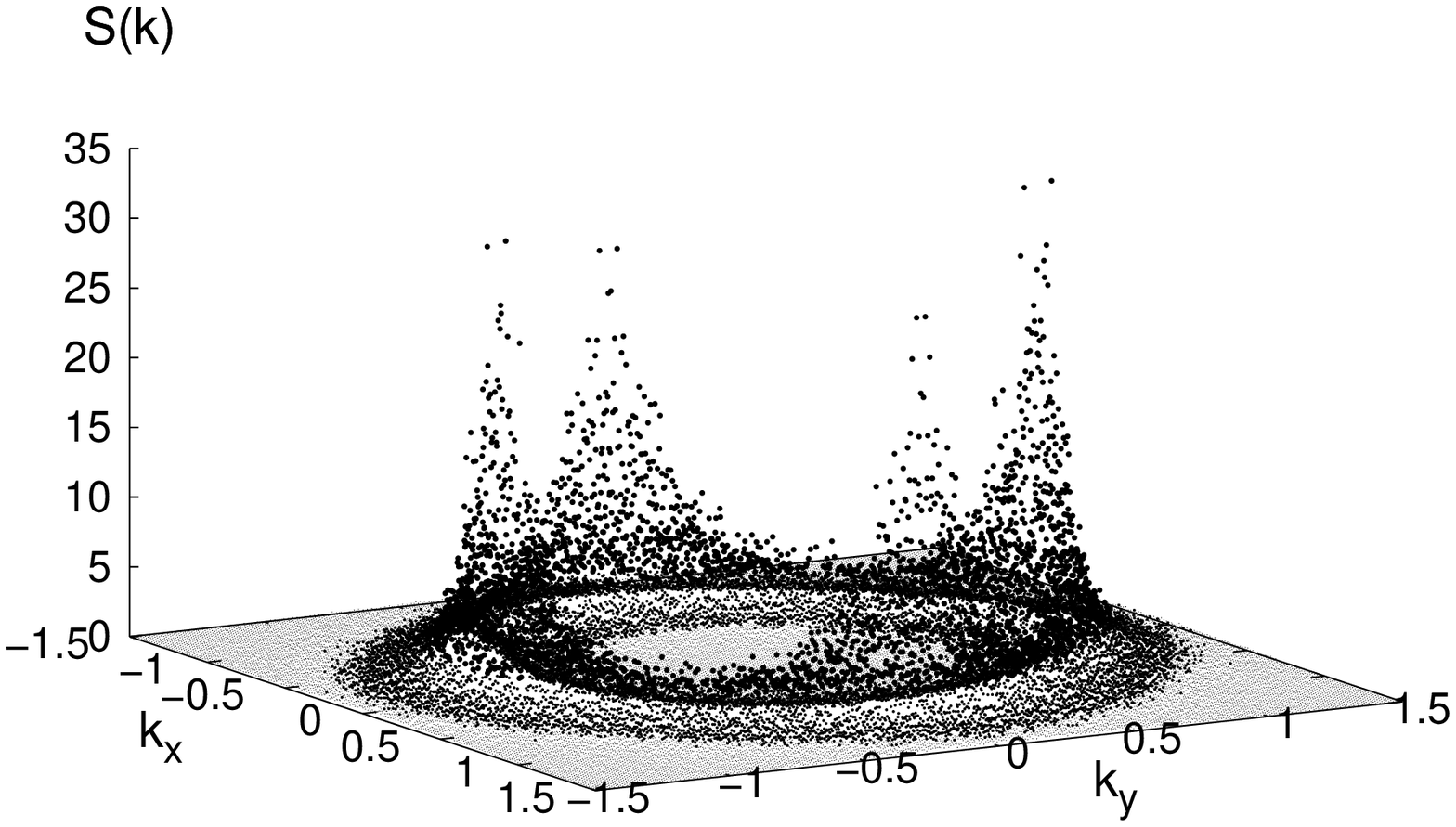}\\
\vs{-1.3}
\includegraphics[width=8cm]{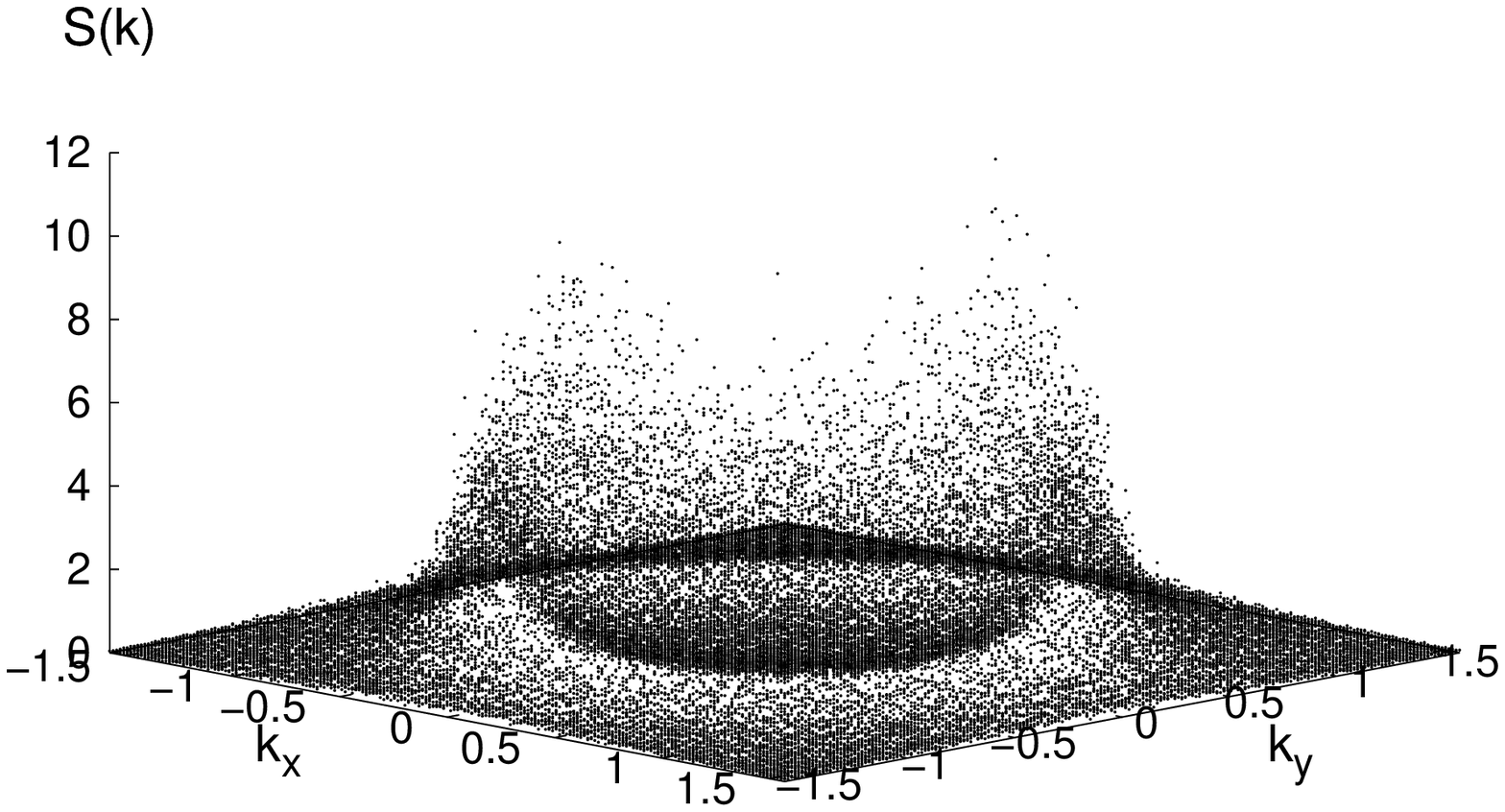}
\caption{Structure factor at $T=0.1$, $0.114$, $0.118$ and $0.14$ for $L=384$.}
\label{sk0.1}
\end{figure}

Finally, we discuss the evolution of the structure factor, eq.(\ref{defSk}), with 
temperature. In Fig.~\ref{sk0.1} we show four characteristic examples around the
transition.
From the first to the second plot it can be seen that the sharp peaks 
characterizing stripe order are replaced by nematic-like peaks, spreaded 
along the ring $|\mb{k}|=k_m$ due to angle fluctuations.
The transition to the disordered phase seems to be 
through an increase of domains of perpendicular orientation of the stripes.
In the structure factor, this is reflected as the growth of two symmetric peaks
around $k\approx k_m$ in the perpendicular direction.
Immediately after the transition, at $T=0.118$, the four symmetrical
peaks are equivalent, as can be seen in the third plot.
As the temperature is further increased, angle fluctuations around 
this two preferential directions smear out the four peaks and 
the system
becomes almost isotropic and the spectral weight of the structure factor
lies on a ring with a weakly tetragonal shape, as shown in Fig.~\ref{sk0.1}.
A typical configuration illustrating the weakly tetragonal symmetry of the
disordered phase just above the transition is shown in Fig.~\ref{defeitos}.

Not far away from the transition, the isotropic phase still presents the exponential-algebraic 
behavior of positions correlations as in the nematic-like regime, 
indicating that it locally resembles the
low temperature phase. The anisotropic feature of the orientational 
correlation function dissapears in this phase, where $C_{nn}$
decays exponentially to zero.



\section{Conclusions}

We have made Langevin simulations at finite temperature of a model for ultrathin ferromagnetic films with
perpendicular anisotropy. These systems have a stripe ground state due to competition between exchange and
dipolar interactions. We have found a phase transition from a high temperature disordered phase
to a low temperature phase with both translational and orientational orders. Below the transition point
isotropy is spontaneously broken, and a smectic-like magnetic structure develops.
Both kinds of ordering seem to
take place at the same $T_c$, in agreement with some previous theoretical predictions. The absence
of an intermediate, purely orientational, phase is probably due to the isotropic nature of the
model Hamiltonian. Other terms, which explicitly break orientational order, may be necessary in
order to have an intermediate nematic phase. This is what emerges from a recent analysis of the nematic
transition in this same model, where an extra term was included and shown to be responsible for the
existence of an intermediate nematic phase~\cite{BaSt2007}. The nature of these terms depend on microscopic 
interactions, namely, on the modelling of the anisotropy contribution. In this respect, more input from experiments
is fundamental.

\begin{figure}[t]
\includegraphics[scale=0.64]{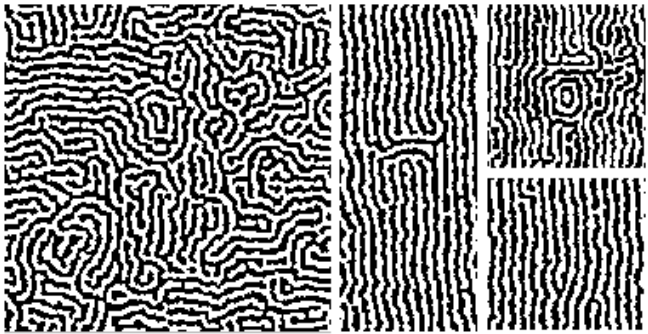}
\caption{The left configuration shows the isotropic phase at $T=0.12$ for $L=192$, presenting a target defect 
close to its center. 
Small target defects are ubiquitous in this phase. At top right we show a large target defect in a portion of 
a $L=480$ configuration 
in an out-of-equilibrium disordering process at $T=0.13$. In the middle a portion of a $L=384$ configuration 
at $T=0.112$ exhibiting a 
large dislocation pair made of disclination pairs. Below, pairs of dislocations at $T=0.104$.}
\label{defeitos}
\end{figure}

We have found that the numerical results show complex magnetic structures, with translational and
orientational correlations decaying algebraically at low temperatures. Translational correlations have different 
exponents for the longitudinal and perpendicular components, relative to the ground state stripe orientation. 
The exponents are also temperature dependent. In the case of orientational correlations, we observe a weak
temperature dependence in the perpendicular direction, and a saturation to a constant value in the 
longitudinal direction. Qualitatively one can say that we observe translational quasi-long-range order and
orientational long range order, at low temperatures. 

As temperature grows it is possible to observe a proliferation of topological defects, with structures
similar to those observed in real systems.
In a narrow region around the transition temperature, both translational and orientational corrrelations
begin to decay exponentially. A power law fit of the data of the orientational correlation length 
and the associated susceptibility in the high temperature region allows us to estimate the critical temperature. This
power law dependence of the orientational correlation length does not agree with the well known KTHNY theory 
of two dimensional melting. We do not know to what extent the predictions from this theory should be
applicable to magnetic two dimensional systems like this. 
Our results are in agreement with a possible scenario for a smectic magnetic system put forward by Abanov 
and coworkers more than 10 years ago~\cite{AbKaPoSa1995}, and with a recent result on the second order
nature of the nematic transition in the same model studied in this work supplemented by another term in the
Hamiltonian which explicitly breaks orientational order~\cite{BaSt2007}. 

We found that the weak anisotropy of the dipolar interaction due to spatial 
discretization
and the short width of the stripes (of three grid points)
has led to a pinning effect that favored the stripes orientation
to be preferentially on the two Cartesian directions.
But looking at the structure factor around the transition,
we found that the fluctuations responsible for disrupting
orientational order are not 
restricted to the two Cartesian directions
and the isotropic nature of the model still manifests.


More quantitative
analytic predictions are clearly needed in order to assess the quality and limitations of our results. 
Also, it
would be extremely interesting to do systematic experimental measurements of structure factors and
correlations as a function of temperature, like the ones done by Pescia and coworkers
~\cite{VaStMaPiPoPe2000,PoVaPe2003,PoVaPe2006}. 
This would allow, between other things,  to elucidate the experimental conditions under which a nematic 
intermediate phase can be present in a magnetic system. 
The similarity between our simulations and experimental images of
perpendicularly magnetized fcc Fe films grown on Cu(100)
\cite{VaStMaPiPoPe2000,PoVaPe2003,PoVaPe2006} is striking. For example, target defects first
observed by Vaterlaus {\it et al.} \cite{VaStMaPiPoPe2000} were observed in our
simulations and are ubiquitous in the disordered phase (see Fig.~\ref{defeitos}).

Finally, the dynamical behavior of these systems also deserve to be studied, because the presence of 
frustration and the emergence of many kinds of topological defects lead to a complex dynamics, with pinning of 
magnetic structures during long time scales, before the final, asymptotic equilibrium state 
sets in~\cite{MuSt2007}.





 The {\em ``Conselho Nacional de Desenvolvimento Cient\'{\i }fico e Tecnol\'{o}gico
CNPq-Brazil''} is aknwoledged for financial support. DAS also wants to aknowledge the Abdus Salam International
Centre for Theoretical Physics for financial support to the {\em Latinamerican Network on Slow Dynamics in
Complex Systems} through grant NET-61.


\end{document}